\documentclass[reprint, aps, prd, showpacs, superscriptaddress, floatfix]{revtex4-2}

\usepackage{graphicx}
\usepackage{amsmath}
\usepackage{amssymb}

\usepackage[caption=false]{subfig}
\usepackage{subcaption}
\usepackage[colorlinks=true, citecolor=blue, urlcolor=blue, linkcolor=blue]{hyperref}

\begin{document}

\title{Exact Analytical Phase Transitions, Horizon Bistability, and Thermodynamic State-Space Representation of Regular Hayward Black Holes}

\author{Jyothipriya M. Shaji}
\email{jyothipriya276@gmail.com}
\affiliation{Independent Researcher}

\author{Jiswin Varghese}
\email{jiswinvarghese@gmail.com}
\affiliation{Department of Physics, Government College, Kattappana, Idukki, Kerala 685508, India}

\author{R. Tharanath}
\email{rtharanath@gmail.com}
\affiliation{Department of Physics, Aquinas College, Edakochi, Ernakulam, Kerala 682010, India}

\author{Sharin B.}
\email{sherinzzsheri@gmail.com}
\affiliation{Independent Researcher}

\date{\today}

\begin{abstract}
 We establish exact analytical thresholds and present a unified thermodynamic state-space representation for the regular Hayward black hole, resolving the full phase structure without reliance on numerical approximations. By evaluating the Hawking temperature, Helmholtz free energy, and heat capacity against the classical Schwarzschild baseline, we derive the exact geometric watershed for the zero-temperature extremal remnant at $r_h = \sqrt3l$, along with the exact Davies critical transition at $r_h = 3l$. Within the intermediate regime $\sqrt{3}l < r_h < 3l$, we uncover a distinct horizon bistability where two distinct horizon radii share identical free energy and temperature profiles. We demonstrate that the Davies singularity acts as a precise thermodynamic divide, separating a locally stable ($C_V > 0$) quantum Small Black Hole branch from an unstable ($C_V < 0$) Large Black Hole branch. Furthermore, we evaluate the exact integrated non-area law entropy at the critical turning point, yielding $S_{turn} = \pi l^2 \left[ \frac{63}{8} + 2\ln(8l^2) \right]$. Finally, we introduce a compact, 3-parameter thermodynamic state-space diagnostic that tracks the continuous evolution of regular black holes from classical thermal evaporation to cold remnant lock, offering a quantitative framework for quantum-gravity phenomenology. 
\end{abstract}

\maketitle

\section{Introduction}\label{sec1}

\subsection{The singularity and Regular Black holes}\label{subsec1}
Classical General Relativity (GR), by Einstein in 1915 \cite{Einstein:1916}  stays as one of the important and highly verified framework in modern theoretical physics. Its predictions, such as gravitational deflection of light, gravitational wave and even  dynamics of expanding spacetime, all stay as proof of how mathematical physics can explain complex structure of our universe. Discoveries such as direct detection of gravitational waves from binary compact object mergers by the LIGO-Virgo-KAGRA collaborations \cite{LIGO:2016} and the sub-horizon scale shadow imaging by the Event Horizon Telescope \cite{EHT:2019}, leave no doubt regarding the existence of astrophysical black holes.
including the direct detection of gravitational waves from binary compact object mergers by the LIGO-Virgo-KAGRA collaborations \cite{LIGO:2016} and the sub-horizon scale shadow imaging by the Event Horizon Telescope \cite{EHT:2019}, leave no doubt regarding the existence of astrophysical black holes. Even though classical general relativity could could explain a lot of such cosmological phenomenon, there was a breaking point for.the theory. The.singularity. According to the foundational singularity theorems  by Penrose and Hawking \cite{Penrose:1965, Hawking:1970}, any gravitational collapse that progresses past an apparent horizon will forces the spacetime geometry to generate a physical singularity. At these points—such as the singularities in the Schwarzschild metric \cite{Schwarzschild:1916}—curvature invariants like the Kretschmann scalar $(K = R^{\mu\nu\alpha\beta}R_{\mu\nu\alpha\beta})$ diverge to infinity. But the classical general relativity failed to these singularities. Which means, proper modifications in GR, are required to overcome such cases.

Even though we do not have a complete quantum theory of gravitation, physicists modeled regular black holes   physicists to study how quantum corrections can help to overcome the above mentioned breakdown points created by singularities. The first of such works were done by Bardeen \cite{Bardeen:1968}, Who proposed the first regular black hole, with the help of a modified metric function. Later, Hayward \cite{Hayward:2006} introduced a physically elegant, alternative regular configuration. Hayward could model black holes with a non-singular core, by modifying the classical Newtonian potential at short distances. At large scales ($r \gg l$, where l is a fundamental quantum-gravitational length scale), the metric behaves as standard Schwarzschild solution smoothly. But at microscopic scale, ${r \to 0}$, the metric transitions into an inner  core with a finite constant curvature, effectively bypassing the geometric singularity.

\subsection{Black Hole Thermodynamics: From Classical to Regular Metrics}\label{subsec2}

The formulation of the four laws of black hole thermodynamics by Bardeen, Carter, and Hawking \cite{Bardeen:1973}, started a journey to combine quantum field theory and  statistical thermodynamics with general relativity. Later Bekenstein proposed that \cite{Bekenstein:1973} the physical entropy of a black hole is directly proportional to its surface area. After that, Hawkings realized that creation of quantum particles near the event horizon causes.the black hole to emit thermal radiations \cite{Hawking:1975}  like real blackbodies at a characteristic temperature $T_H$.

Classical, singular black holes shows an inverse relationship between mass and temperature ($T_H \propto \frac{1}{M}$), this causes a negative specific hear capacity for the black hole which means it becomes hotter as radiates away its mass. this makes the black hole unstable and eventually its whole mass is radiates away and disappears. This complete disappearance introduces the notorious black hole information paradox \cite{Hawking:1976}, 
In the case of a non-singular Hayward black hole, this Thermodynamic behavior is changed. Since the metric tensor includes a non-polynomial curvature structure, the standard assumptions of horizon behavior and area laws cannot be directly applicable here. Recent studies have shown that regular horizons exhibit unique phase behaviors and thermodynamic boundaries \cite{Kumar:2026}. so it is important to understand, how the elimination of central singularity modifies observable parameters, such as Hawking temperature and specific heat. It helps us to understand, how changing the microscopic space time geometry can naturally stop thermal runaway and how is they are distinguished from classical black holes with singularities.

\subsection{Outline of this work}\label{subsec3}
Through this paper, we are systematically evaluate the thermodynamic phase structure, local stability configurations, and global potentials of the vacuum Hayward regular black hole by comparing its behavior directly against the classical Schwarzchild baseline. The paper is structured as follows:
In Section 2, we will introduce the geometric setup of the Hayward metric function, then derive an exact analytical mass function.
Section 3 provides the core mathematical framework of this study, where we derive the modified Hawking temperature, specific heat capacity, and Helmholtz free energy. Since this is a black hole with  non-polynomial curvature of the regular core, we integrate the first law of black hole thermodynamics to derive the mathematically exact, non-area law entropy profile.
In Section 4, we perform, a side-by-side comparative thermodynamic analysis. Sine the Hayward black hole acts as classical Schwarzschild black hole as  $l \to 0$, we will compare the regular Hayward properties directly against the classical Schwarzchild limits, in order to understand their stability profiles, locate the second-order thermodynamic phase transitions at the Davies point, and investigate the emergence of a frozen, zero-temperature remnant phase.
Finally, in Section 5, we summarize our primary conclusions and discuss the broader cosmological implications of these regular geometries regarding the black hole information paradox and the viability of stable remnants as primordial dark matter candidates.

\section{The Vacuum Hayward Black Hole and Horizon Framework}\label{sec2}
\subsection{Metric Construction and Geometric Boundaries}\label{subsec4}

To investigate the thermodynamic modifications by  a non-singular spacetime core, we consider the regular, spherically symmetric vacuum Hayward black hole in four dimensions. The spacetime geometry is described by the line element:
\begin{equation}
ds^{2} = -f(r)dt^{2} + \frac{1}{f(r)}dr^{2} + r^{2}(d\theta^{2} + \sin^{2}\theta d\phi^{2})
\label{eq:hayward_metric}
\end{equation}
where the metric function $f(r)$ is defined as \cite{Hayward:2006}:
\begin{equation}
f(r) = 1 - \frac{2Mr^{2}}{r^{3} + 2Ml^{2}}
\label{eq:metric_func}
\end{equation}
Here, M is the  mass parameter of the configuration, and $l$ is a parameter introducing a fundamental length scale tracking a quantum-gravitational core cutoff.
As we can see, The  equation \eqref{eq:metric_func} can cure the the classical singularity problem via two distinct limiting regimes:
\begin{itemize}
    \item \textbf{The Macroscopic Large-Scale Limit $(r\gg l)$}: When the radial coordinate     significantly greater than $l$, the denominator becomes, $ r^3 + 2Ml^2 \approx r^3 $.     So the metric function  reduces to:
        \begin{equation}
            f(r) \approx 1 - \frac{2M}{r} + \mathcal{O}\left(\frac{l^2}{r^3}\right)
        \end{equation}
        This means, at large distances, the the Hayward solution smoothly behaves as the classical vacuum Schwarzschild solution \cite{Schwarzschild:1916}.
    \item \textbf{The Microscopic Small-Scale Limit $(r \to 0)$}: As we reach the center of the geometry, the $r^3$ term inside the denominator vanishes relative to the constant scale term $(r^3 \ll 2Ml^2)$. Taylor expanding the metric function around the origin yields:
    \begin{equation}
        f(r) \approx 1 - \frac{r^2}{l^2} + \mathcal{O}(r^3)
    \end{equation}
    At this region, it acts as a regular de Sitter spacetime background \cite{Einstein:1916} characterized by an effective cosmological constant $\Lambda_{\text{eff}} = 3/l^2$. 
\end{itemize}

\subsection{Causal Horizon Framework and the Exact Mass Function}\label{subsec5}
At the horizon, the temporal metric component vanishes $(f(r) = 0)$. From Equation \eqref{eq:metric_func}, setting $f(r_h) = 0$, gives:
\begin{equation}
1 - \frac{2Mr_h^{2}}{r_h^{3} + 2Ml^{2}} = 0 \implies r_h^{3} + 2Ml^{2} = 2Mr_h^{2}
\label{eq:horizon_algebraic}
\end{equation}
where $r_h$ is the horizon radius.
The equation \eqref{eq:horizon_algebraic} simplifies to:
\begin{equation}
2M(r_h^2 - l^2) = r_h^3
\end{equation}
Which gives the exact analytical mass function of the horizon radius:
\begin{equation}
M(r_h) = \frac{r_h^{3}}{2(r_h^{2} - l^{2})}
\label{eq:analytical_mass}
\end{equation}
An algebraic analysis of Equation \eqref{eq:analytical_mass} shows that the mass function depends entirely on the presence of the regularization scale l:

\begin{itemize}
    \item Physical Horizon Convergence: For a positive mass parameter (M > 0), any valid horizon configuration must strictly greater than $l$ $(r_h > l)$. If the horizon value is less than l, then the equation of mass becomes negative which is nonphysical. 
    \item The Extremal Threshold: we can locate the minimum mass threshold  required to hold an open horizon by taking derivative of equation \eqref{eq:analytical_mass} with respect to $r_h$ :
\begin{equation}
\begin{aligned}
\frac{dM}{dr_h} = \frac{3r_h^2 \cdot 2(r_h^2 - l^2) - r_h^3 \cdot 4r_h}{4(r_h^2 - l^2)^2} \\ \\
        = \frac{6r_h^4 - 6r_h^2l^2 - 4r_h^4}{4(r_h^2 - l^2)^2} \\ \\
        = \frac{r_h^2(r_h^2 - 3l^2)}{2(r_h^2 - l^2)^2}
\label{eq:mass_derivative}
\end{aligned}
\end{equation}
Setting $\frac{dM}{dr_h} =0$ will show us a  stable critical turning point :
\begin{equation}
r_{\text{ext}} = \sqrt{3}l
\label{eq:r_extremal}
\end{equation}
We can find a critical minimum mass $(M_{\text{crit}})$configuration by substituting the extremal radius back into the global mass relation \eqref{eq:analytical_mass},    below which an event horizon cannot form classically:
\begin{equation}
M_{\text{crit}} = M(r_{\text{ext}}) = \frac{(\sqrt{3}l)^3}{2((\sqrt{3}l)^2 - l^2)} = \frac{3\sqrt{3}l^3}{4l^2} = \frac{3\sqrt{3}}{4}l
\label{eq:m_critical}
\end{equation}
\end{itemize}
For any mass parameter $M > M_{crit}$ , we can see two different real positive roots: one can be interpreted as an outer event horizon $(r_+)$ and the other one can be interpreted as an inner Cauchy horizon $(r_-)$. When $M = M_{crit}$, these surfaces will merge into a single horizon $(r_+ = r_- = \sqrt{3}l)$, signaling an extremal regular remnant phase.. Similarly , if $M < M_{crit}$, no real solutions exist for $f(r_h)=0$, and the system transitions into a stable, ultra-compact horizon-less configuration.

\section{Analytical Derivation of Thermodynamic Variables }\label{sec3}
\subsection{Modified Hawking Temperature Profile }\label{subsec6}

 The Hawking temperature $T_H$ is  linked to the geometric surface gravity $\kappa$ evaluated at the outer event horizon ($r_h$) via the relation $T_H = \kappa / 2\pi$. For a static, spherically symmetric spacetime, this reduces to evaluating the first radial derivative of the metric function at the horizon radius:

\begin{equation}
    T_{H} = \frac{f'(r_h)}{4\pi}
    \label{eq:temp_base}
\end{equation}

Differentiating the metric function Equation \eqref{eq:metric_func} with respect to $r$:
\begin{equation}
\begin{aligned}
f'(r) = -\frac{\partial}{\partial r}\left[\frac{2Mr^{2}}{r^{3} + 2Ml^{2}}\right] \\ \\= -\frac{4Mr(r^{3} + 2Ml^{2}) - 2Mr^{2}(3r^{2})}{(r^{3} + 2Ml^{2})^{2}}\\  \\= \frac{2Mr^{4} - 8M^{2}l^{2}r}{(r^{3} + 2Ml^{2})^{2}}
\end{aligned}
\end{equation}
At the event horizon boundary, we substitute the exact horizon constraint $r_h^3 + 2Ml^2 = 2Mr_h^2$  from equation \eqref{eq:horizon_algebraic} into the denominator, which drastically simplifies the expression to:
\begin{equation}
\begin{aligned}
f'(r_{h}) = \frac{2Mr_h^{4} - 8M^{2}l^{2}r_h}{(2Mr_h^2)^{2}} \\\\
= \frac{2Mr_h^4 - 8M^2l^2r_h}{4M^2r_h^4}\\\\
= \frac{r_h^3 - 4Ml^2}{2Mr_h^3}
\label{eq:f_prime_simp}
\end{aligned}
\end{equation}
To express the temperature purely as a function of the horizon radius, we eliminate the mass parameter M by inserting the exact analytical mass relation $M(r_h) = \frac{r_h^3}{2(r_h^2 - l^2)}  $ from Equation \eqref{eq:analytical_mass} into Equation \eqref{eq:f_prime_simp}:
\begin{equation}
\begin{aligned}
f'(r_{h}) = \frac{r_h^3 - 4\left(\frac{r_h^3}{2(r_h^2 - l^2)}\right)l^2}{2\left(\frac{r_h^3}{2(r_h^2 - l^2)}\right)r_h^3} \\\\ = \frac{r_h^3 - \frac{2r_h^3l^2}{r_h^2 - l^2}}{\frac{r_h^6}{r_h^2 - l^2}}\\\\ = \frac{r_h^3(r_h^2 - l^2) - 2r_h^3l^2}{r_h^6}\\\\ = \frac{r_h^2 - 3l^2}{r_h^3}
\end{aligned}
\end{equation}
Substituting this geometric result back into the temperature equation \eqref{eq:temp_base} , we get  the exact analytical Hawking temperature function for the vacuum Hayward metric:
\begin{equation}
T_{H}(r_{h}) = \frac{r_{h}^{2} - 3l^{2}}{4\pi r_{h}^{3}}
\label{eq:analytical_temp}
\end{equation}
A simple evaluation of Equation \eqref{eq:analytical_temp} shows the role of the regular scale $l$. In the macroscopic limit $(l \to 0$  or $r_h \gg l)$, the temperature becomes the classical Schwarzschild limit $T_H$ $\approx 1/(4\pi r_h)$, which diverges to infinity as the radius shrinks. Also we will see, inside the Hayward framework, the temperature reaches a finite maximum and then drops continuously to absolute zero $(T_H = 0)$ exactly at the critical extremal radius $r_h = \sqrt{3}l$.

\subsection{Exact First-Law Integration of Horizon Entropy}\label{subsec7}
Since the modified Hayward metric includes a non-polynomial curvature structure, we cannot use the standard Bekenstein area law $(S = A/4 = \pi r_h^2)$  \cite{Kumar:2026}. But we can find the mathematically true entropy profile from the  first law of black hole thermodynamics, $dS = \frac{dM}{T_H}$:
\begin{equation}
dS = \frac{1}{T_{H}}\left(\frac{\partial M}{\partial r_{h}}\right)dr_{h}
\label{eq:entropy_first_law}
\end{equation}
From our previous derivation of the mass gradient in Equation \eqref{eq:mass_derivative}, we have 
\begin{equation}
\frac{\partial M}{\partial r_h} = \frac{r_h^2(r_h^2 - 3l^2)}{2(r_h^2 - l^2)^2}
\end{equation}
Substituting this and Hawking temperature \eqref{eq:analytical_temp} into Equation \eqref{eq:entropy_first_law} :
\begin{equation}
\begin{aligned}
dS = \left(\frac{4\pi r_{h}^{3}}{r_{h}^{2} - 3l^{2}}\right) \cdot \left(\frac{r_{h}^{2}(r_{h}^{2} - 3l^{2})}{2(r_{h}^{2} - l^{2})^{2}}\right) dr_{h} \\\\
= \frac{2\pi r_{h}^{5}}{(r_{h}^{2} - l^{2})^{2}} dr_{h}
\label{eq:entropy_diff}
\end{aligned}
\end{equation}
and from this equation we can find entropy, 
\begin{equation}  
    S(r_h) = \int \frac{2\pi r_h^5}{(r_h^2 - l^2)^2} dr_h
\end{equation}
let  $u = r_h^2 - l^2$, which implies $r_h^2 = u + l^2$ and $du = 2r_h dr_h$. \\
The integral transforms as follows:
\begin{equation}
\begin{aligned}
        S(u) = \pi \int \frac{(u + l^2)^2}{u^2} du \\\\ 
             = \pi \int \frac{u^2 + 2l^2u + l^4}{u^2} du  \\\\
             = \pi \int \left( 1 + \frac{2l^2}{u} + \frac{l^4}{u^2} \right) du
\end{aligned}
\end{equation}
Integrating this term:
\begin{equation}
S(u) = \pi \left[ u + 2l^2\ln(u) - \frac{l^4}{u} \right] + S_0
\label{eq:entropy_u}
\end{equation}
where $S_0$ is a constant of integration. 
substituting for u will give and exact global entropy expression for the regular Hayward black hole:
\begin{equation}
S(r_{h}) = \pi \left[ (r_{h}^{2} - l^{2}) + 2l^{2}\ln\left({r_{h}^{2} - l^{2}}\right) - \frac{l^{4}}{r_{h}^{2} - l^{2}} \right]
\label{eq:analytical_entropy}
\end{equation}

We can see in \eqref{eq:analytical_entropy} as $l \to 0$ reveals that the  logarithmic and rational correction terms vanish completely, returning back to the classical Bekenstein area formula $S = \pi r_h^2$.

Evaluating the global entropy expression \eqref{eq:analytical_entropy} at the Davies critical transition boundary ($r_h = 3l$), which marks the thermodynamic divide between the locally stable ($C_V > 0$) and unstable ($C_V < 0$) branches, yields the exact non-area law critical entropy:
\begin{equation}
S_{\text{turn}} = \pi l^2 \left[ \frac{63}{8} + 2\ln(8l^2) \right]
\label{eq:S_turn}
\end{equation}
Equation \eqref{eq:S_turn} demonstrates that the logarithmic quantum-geometric correction dominates the microstate count near the critical watershed, providing an explicit upper bound before the system enters the cold remnant phase.

\subsection{Specific Heat Capacity and Stability Formulations}\label{subsec8}
Local stability of a black hole can be analyzed by studying its Heat capacity:
\begin{equation}
C = \frac{\partial M}{\partial T_{H}} = \frac{\partial M / \partial r_{h}}{\partial T_{H} / \partial r_{h}}
\label{eq:heat_chain_rule}
\end{equation}
From equation(\eqref{eq:analytical_temp} ):
\begin{equation}
    \begin{aligned}
        \frac{\partial T_{H}}{\partial r_{h}} = \frac{\partial}{\partial r_h}\left[\frac{r_h^2 - 3l^2}{4\pi r_h^3}\right] \\\\
        = \frac{2r_{h}(4\pi r_{h}^{3}) - (r_{h}^{2} - 3l^{2})(12\pi r_{h}^{2})}{(4\pi r_{h}^{3})^{2}}\\\\
        = \frac{8\pi r_h^4 - 12\pi r_h^4 + 36\pi l^2r_h^2}{16\pi^2 r_h^6} \\\\
        = \frac{9l^{2} - r_{h}^{2}}{4\pi r_{h}^{4}}
        \label{eq:temp_derivative}
    \end{aligned}
\end{equation}
By combining equation \eqref{eq:temp_derivative}  with equation (\eqref{eq:mass_derivative}), equation (\eqref{eq:heat_chain_rule}),  the heat capacity relation:
\begin{equation}
\begin{aligned}
        C(r_{h}) = \frac{r_{h}^{2}(r_{h}^{2} - 3l^{2})}{2(r_{h}^{2} - l^{2})^{2}} \cdot \left( \frac{4\pi r_{h}^{4}}{9l^{2} - r_{h}^{2}} \right)\\\\
        = \frac{2\pi r_{h}^{6}(r_{h}^{2} - 3l^{2})}{(r_{h}^{2} - l^{2})^{2}(9l^{2} - r_{h}^{2})}
\label{eq:analytical_heat}
\end{aligned}
\end{equation}
Equation \eqref{eq:analytical_heat} governs our stability analysis. 
\subsection{ Helmholtz Free Energy}
To study about stability and possible phase transmission  of the black hole,  we can derive the  Helmholtz free energy potential by mapping the black hole mass $M$ as the total internal energy of the spacetime configuration, the Legendre transformation follows the standard textbook framework \cite{Callen:1985},  $F = M - T_H S$. Substituting Equations \eqref{eq:analytical_mass}, \eqref{eq:analytical_temp}, and \eqref{eq:analytical_entropy} into this transformation gives:
\begin{equation}
\begin{aligned}
        F(r_{h}) = \frac{r_{h}^{3}}{2(r_{h}^{2} - l^{2})} - \left[ \frac{r_{h}^{2} - 3l^{2}}{4\pi r_{h}^{3}} \right] \\  \cdot \pi \left[ (r_{h}^{2} - l^{2}) + 2l^{2}\ln(r_{h}^{2} - l^{2}) - \frac{l^{4}}{r_{h}^{2} - l^{2}} \right]
\label{eq:analytical_free_energy}
\end{aligned}
\end{equation}

\section{Result and discussion}\label{sunsec}\label{sec4}

In order to understand the effect of curing the central singularity,  we perform a side-by-side comparative analysis of the regular Hayward black hole against the classical Schwarzschild baseline. since the length parameter is important in avoiding the singularity, it is a good way to analyze their dependence on the behavior of the black hole too, so we used 3 values for l, $-0.5,1,0.5$. the results are given in the following section.  even though we used different values for $l$, all the graphs looks similar for each properties. 

\subsection{Mass vs. Horizon radius/Entropy Layouts}
In the macroscopic limit $(r_h \gg l)$, the mass function of the regular Hayward black hole  behaves similar to that of the standard linear Schwarzschild relation:
\begin{equation}
\lim_{l \to 0} M(r_h) = \frac{r_h}{2}
\end{equation}
However, plotting the full mass profile against either the horizon radius $r_h$ or the exact integrated entropy S uncovers a distinct geometric divergence between the two frameworks.

\begin{figure*}[t]
    \centering
    \subfloat[M vs r\label{fig:m_vs_r}]{%
        \includegraphics[width=0.48\textwidth]{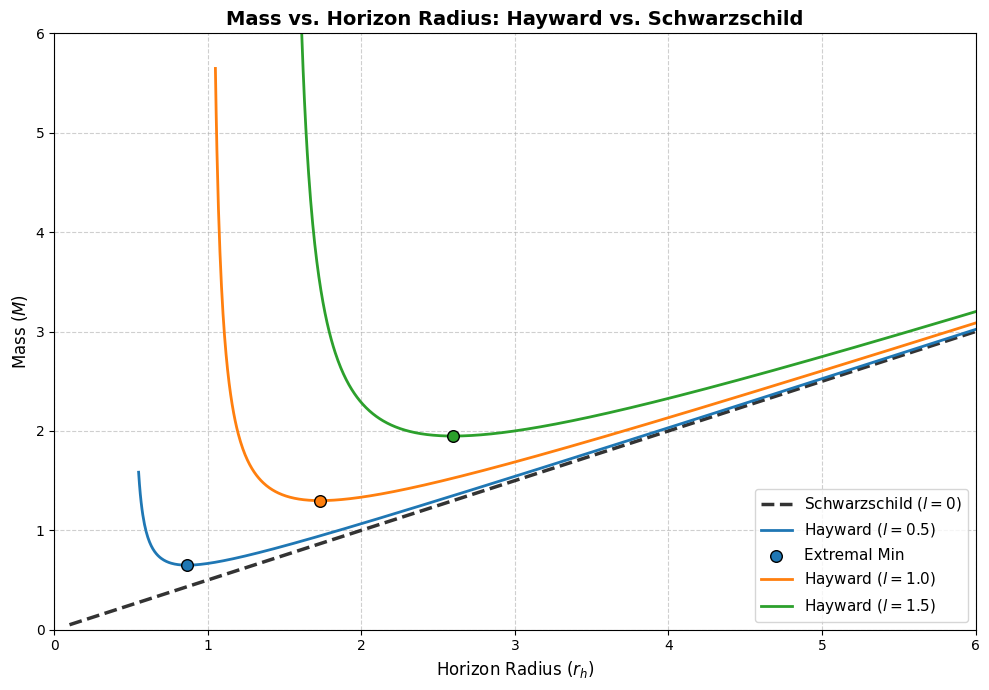}%
    }%
    \hfill
    \subfloat[M vs S\label{fig:m_vs_s}]{%
        \includegraphics[width=0.48\textwidth]{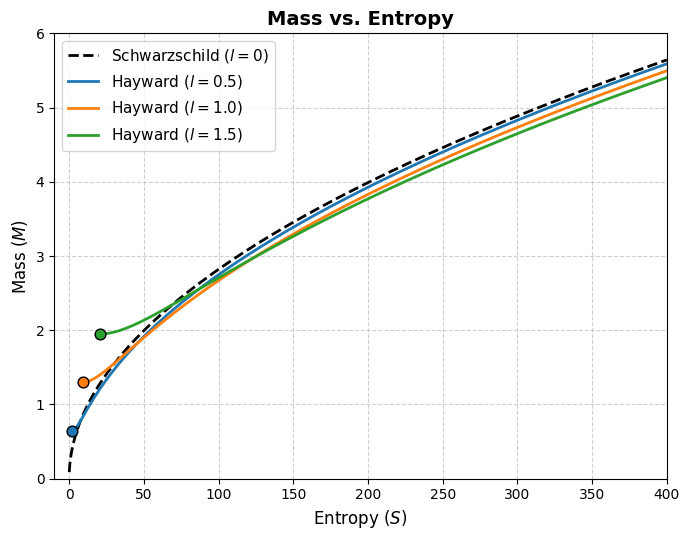}%
    }%
    \caption{The behavior of mass of the Hayward black hole is shown here. Figure~\ref{fig:m_vs_r} shows how mass of the black hole depends on its radius and figure~\ref{fig:m_vs_s} shows how it changes with entropy. The dashed lines show the behavior of a Schwarzschild black hole. The main point to note here is how the Hayward black hole avoids the singularity.}
    \label{fig:M_set}
\end{figure*}

For the classical Schwarzschild case, the mass decreases monotonically as the horizon contracts, shrinking all the way to a zero-mass, zero-radius point $(M=0$ at $r_h=0)$. In contrast, the Hayward mass profile displays a distinctive "hook-shaped" minimum threshold. As mass is lost during evaporation, the horizon contracts until it encounters the critical turning point isolated in Equation \eqref{eq:r_extremal} at $r_{\text{ext}} = \sqrt{3}l$. At this point, the mass reaches its minimum allowable threshold, $M_{\text{crit}} = \frac{3\sqrt{3}}{4}l$.
Below this turning point, any further reduction in horizon radius requires an infinite injection of mass $(M \to \infty$ as $r_h \to l^+)$. This asymptotic barrier stems from the core de Sitter geometry, which creates an effective repulsive quantum-gravitational force at ultra-short distances, halting complete gravitational collapse.

\textbf{Mass versus Horizon Radius $r_h$}\\
Since the Hayward regular black holes could avoid singularity for smaller radius, where the regularization length,  $l$ dominates, and becomes similar to Schwarzschild black holes in higher radius, the same physical situation can be observed here. in the macroscopic limit, which means $r>>l$, the solid lines merges with dashed lines. this shows that at higher radius the Hayward blackhole is similar to Schwarzschild black hole. so we can say that the quantum corrections becomes indistinguishable from classical gravity. 
the "regularization" region shows the interesting features of Hayward black hole. here as the radius shrinks, the  mass of Schwarzschild black hole approaches zero. but in the case of Hayward metric, it prevents the object to reach a zero value. which means, as the object collapses, mass required to create a horizon increases drastically, As mass is lost during evaporation, the horizon contracts until it encounters the critical turning point isolated in Equation \eqref{eq:r_extremal} at $r_{\text{ext}} = \sqrt{3}l$. At this point, the mass reaches its minimum allowable threshold, $M_{\text{crit}} = \frac{3\sqrt{3}}{4}l$.  which means the mass required to create such a small radius is affected by a repulsive nature created by the quantum correction. this stops the horizon from collapsing below l.
another interesting feature that we note here is a minimum value that is required to create a horizon. as the horizon increases from zero, we can see the mass required decreases from very high value as it reaches a minimum value, then it starts to increase. this behavior can be interpreted as, if a star of any huge gravitating body collapses, if the total mass of the object is below this minimum value, subjected to the given value of regularization length, the object is unable to create an event horizon instead it collapses into a regular, horizon-less compact object. 
another interesting feature is, for any radius, above the corresponding value of minimum mass point, if we extend a straight line towards the mass value, it touches the plot in two points. which means for same mass in that region, there are two possible event horizons. which means the geometry possesses a dual event horizon structure. an outer event horizon and an inner event horizon. and as the black hole evaporates its mass, the outer event horizon shrinks and the inner event horizon expands and reaches at the extremal limit. which means it reaches a stable point. the same can be verified by observing temperature versus radius graph.\\

\textbf{Mass versus entropy}\\
 the figure \ref{fig:m_vs_s} The graph follows  a hook shaped structure for different values of l. Hence let us split this region into 3 zones;\\

\textbf{Zone 1: the linear upper branch (large entropy)} 
As we move to higher entropy values, mass become approximately increases with entropy and its behavior is similar to that of Schawrzschild black hole.
In this region, mass is directly proportional to $\sqrt{S}$ because $r>>l$.  this region act like the ordinary Schwarzschild black holes\\

\textbf{Zone 2: the critical turning point (M is minimum) }
When we study the curve from right to left, that is, the black hole is evaporating and closing entropy, the region where the curve gives a sharp minimum value .\\ Since there are no mass to support entropy lower than this, we can assume that, below this mass, gravitational collapse cannot produce a black hole. Instead the collapsing object remains as a regular compact object without an event horizon. so we can say, from the graph, this value of mass separate ordinary black hole from other compact objects without and event horizon.\\

\textbf{Zone 3: small entropy region}
instead of mass goes to zero, as Schawrzschild black hole, Smaller entropy values do not support mass. If we try to study this lower entropy region with mass vs $r_h$ graph, we can safely assume that the quantum correction in Hayward metric is producing a repulsive mechanism.
\subsection{Temperature vs. Horizon radius/Entropy}
This graphs are providing information about temperature of the Black hole as well as its evaporation. Similar to mass, Critical turning points can be observed here also. 

\begin{figure*}[t]
    \centering
    \subfloat[T vs r\label{fig:T_vs_r}]{%
        \includegraphics[width=0.48\textwidth]{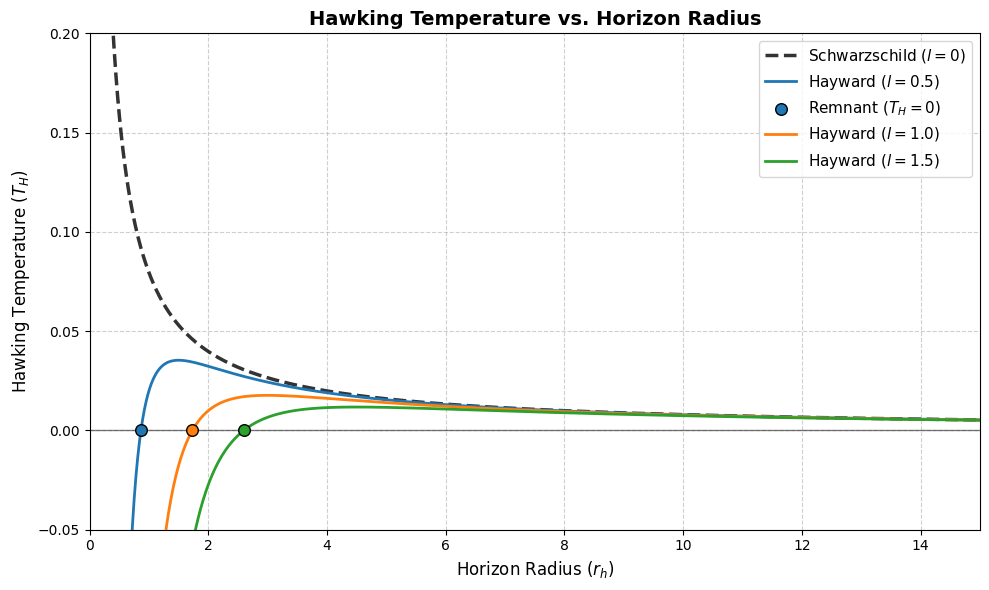}%
    }%
    \hfill
    \subfloat[T vs S\label{fig:T_vs_s}]{%
        \includegraphics[width=0.48\textwidth]{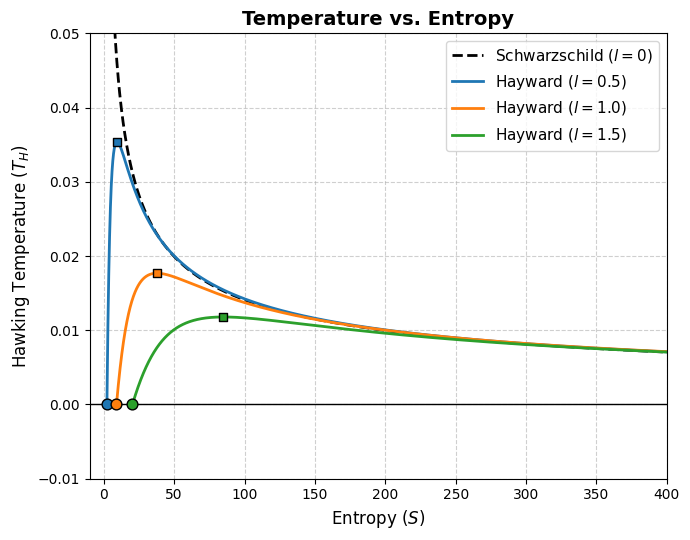}%
    }%
    \caption{The figure shows the change in temperature of the Hayward black hole with radius \ref{fig:T_vs_r} shows its behavior with temperature while \ref{fig:T_vs_s} shows that with entropy. both graphs shows similar behavior but it should be noted that, unlike Schwarzschild black hole, Hayward black hole shows an opposite trend on lower values and it have a maximum possible temperature. }
    \label{fig:T_set}
\end{figure*}

\textbf{Temperature versus horizon radius}
Similar to mass - horizon study, here also the black hole behaves as Schwarzschild black hole at larger radii.  It is shown in \ref{fig:T_vs_r}. but the interesting part of the graph is, its maximum possible temperature. the temperature increases and reaches a maximum value then it decreases. it shows a local maxima. also we can observe that the solid line crosses the radius axis from the negative temperature region, which means for that particular radius the Hawking radiation stops which means it leaves a cold, stable core. also when compare it with mass - radius graph, the minima in that graph is near the maxima of the temperature - radius graph.  \\These are the same values of radius, where the mass horizon graph shown the minima. if we look at the Schwarzschild black hole, its temperature increases as we move from larger radii to smaller radii. \\
The effect of the  strength of the quantum correction should be noted here. Larger value of $l$ pushes the maximum value of temperature towards higher radius values. so as the radius value which touches the zero temperature. 

\textbf{Temperature versus Entropy}

The figure \ref{fig:T_vs_s} explains, how cold the black hole get as it losses  the entropy. Same as other quantities, the black holes behave as Schwarzschild black hole. That is, a slight increase in temperature while entropy decrease. \\
let's look at the graph from right to left, in order to understand the evolution of the black hole. As the black hole is large, it  behaves as a Schwarzschild black hole. and as it evaporates, its radius as well as entropy is decreased. As it continues its evaporation, its temperature increases. but when the entropy reaches certain values, the temperature reaches its maximum value then further shrinking will not increase the temperature of the black hole, instead it begins to cool. but at this point, the Schwarzschild black hole will be heating indefinitely. \\
finally entropy reaches int minimum value, here the Hawking temperature becomes zero. This vanishing of temperature can be assumed as the Hawking radiation was stopped there.   leaving behind a cold external remnant. \cite{Chen:2015} \\
also we can see, 
As evaporation continues, the Hawking temperature reaches a maximum. This maximum marks the onset of strong quantum corrections. From this point onward, the evaporation process changes qualitatively: instead of heating indefinitely, the black hole begins to cool. \\
Another interesting feature we can identify is, as the value of regularization length $l$  increases, the maximum Hawking  temperature decreases. which means, Stronger the quantum effect, lower the maximum Hawking temperature. Also, stronger quantum effects, suppresses the Hawking temperature earlier.  
As we saw in the case of Mass, here also two entropy values will support same temperature.

\subsection{Specific heat vs. Horizon radius/Entropy}

Heat capacity shows the stability of the black hole. a negative heat capacity shows an unstable black hole and it gets hotter as it radiates while a positive heat capacity shows a stable one.

\begin{figure*}[t]
    \centering
    \subfloat[C vs r\label{fig:C_vs_r}]{%
        \includegraphics[width=0.48\textwidth]{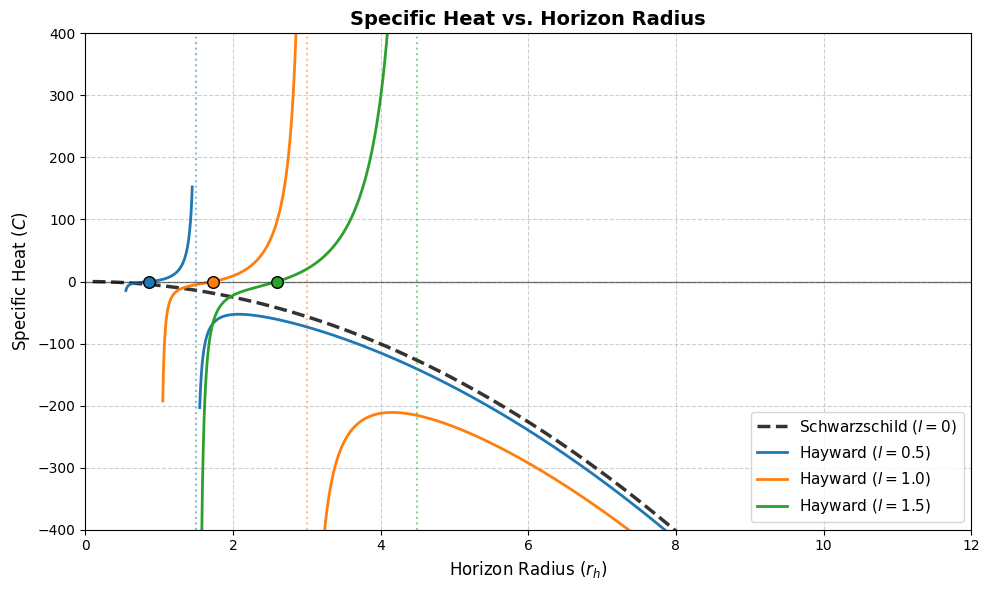}%
    }%
    \hfill
    \subfloat[C vs S\label{fig:C_vs_s}]{%
        \includegraphics[width=0.48\textwidth]{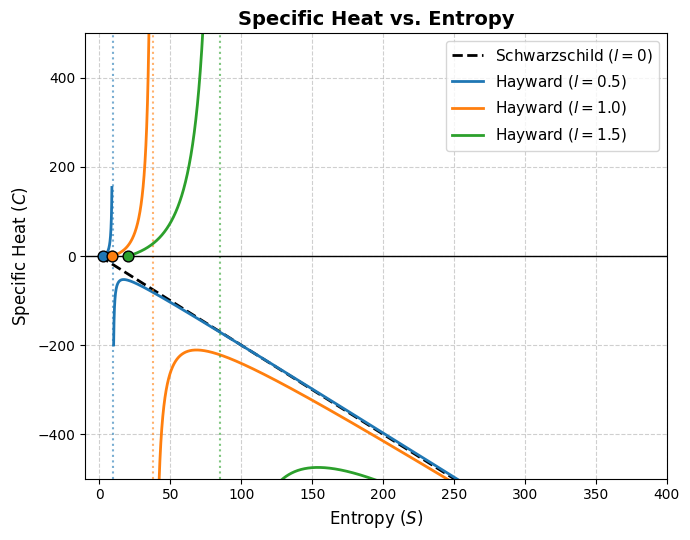}%
    }%
    \caption{This figure shows change in specific heat as radius changes (\ref{fig:C_vs_r} and as entropy changes \ref{fig:C_vs_r}. both graphs shows information about stability of black hole. also we can see it mimics the behavior of Schwarzschild black hole in higher radii. }
    \label{fig:C_set}
\end{figure*}

Since the Hawking temperature is inversely proportional to Mass,  as the mass increases, the temperature decreases . To understand specific heat, using \eqref{eq:analytical_heat}, we can say, losing that mass heats up the black hole and it radiates faster. Which makes the black hole unstable. 

\textbf{Specific heat versus radius }\\
As we said, for a stable black hole, the information of its specific heat brings information of its stability. we can see that in the figure \ref{fig:C_vs_r}. if we move from right to left, the direction where the radius decreases, we can say the black hole radiates and losses its mass so as its radius, Its temperature increases and it becomes hotter. so its specific heat stays negative for larger radii and they are unstable in that range. we can see the same behavior in the case of Schwarzschild black hole. 

Moving from  right to left, the black hole shrinks but at some point, the curve reaches a discontinuity. which means there could be a phase transition. Left of the discontinuity, the specific heat is positive and right of the discontinuity specific heat is negative. this point of disconnected is the same region where we could see a local minima in the case of mass vs horizon graph and a local maxima in the case of temperature vs r graph. so we can assume  these points are critical points. 

We can see that left of this point, the black hole is thermodynamically stable because of the quantum effects and the specific heat reaches a maximum. Then as it shrinks more, its specific heat reduces, it continues to lose mass and it cools down then reaches zero. 
 
 \textbf{Specific Heat versus Entropy}\\
At larger entropy the black hole behaves similar to Schwarzschild black hole. Specific heat is negative, indicating that, it radiates more losses more energy. So it is unstable.

As the entropy decreases, the specific heat diverges at a critical entropy. This divergence signals a possible second-order phase transition, commonly referred to as the Davies transition \cite{Davies:1977}. After this critical point, we can see that, unlike Schwarzchild black hole, the specific heat stays positive, which indicates that the black hole is at a thermodynamically stable quantum phase.  for smaller entropy we can see that the specific heat becomes zero, indicating that the black hole stops radiating energy. which means,  the Hayward black hole naturally evolves toward a stable remnant through this phase transition.
It is clear that regularization parameter have strong effect on the phase transition. Larger the $l$  , the earlier the phase transition point and earlier the black hole attains stability.    
Increasing the regularization parameter shifts the transition to larger entropy, indicating that stronger quantum corrections stabilize the black hole at an earlier stage of its evolution. 

\subsection{Helmholtz Free energy vs. Horizon radius/Entropy}
The Helmholtz free energy can be interpreted as the  amount of energy available for the system to perform thermodynamic works. Lower free energy , means the system is in a Thermodynamically favorable configuration,

\begin{figure*}[t]
    \centering
    \subfloat[C vs r\label{fig:F_vs_r}]{%
        \includegraphics[width=0.48\textwidth]{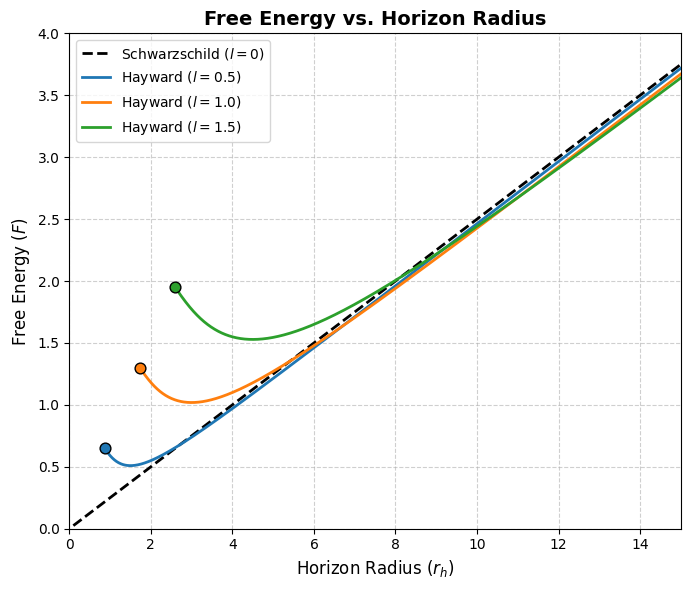}%
    }%
    \hfill
    \subfloat[C vs S\label{fig:F_vs_s}]{%
        \includegraphics[width=0.48\textwidth]{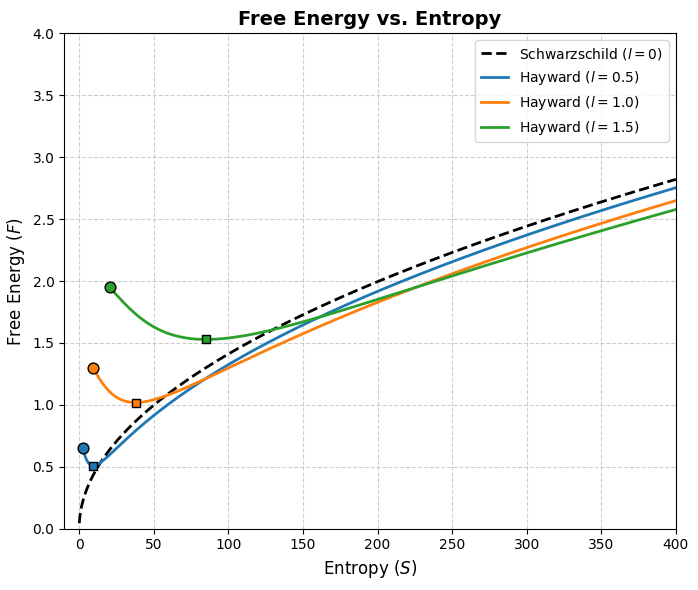}%
    }%
     \caption{This figure shows change in specific heat as radius changes (\ref{fig:C_vs_r} and as entropy changes \ref{fig:C_vs_r}. both graphs shows information about stability of black hole. also we can see it mimics the behavior of Schwarzschild black hole in higher radii. }
    \label{fig:C_set}
\end{figure*}

\textbf{Helmholtz free energy vs. Horizon radius\\}
For large horizon radii $(r_h \gg l)$, all Hayward curves approach the Schwarzschild solution. This means, that like all other parameters, the effects of the regularization parameter become negligible at large scales. So for larger radii,  the thermodynamic behavior of large Hayward black holes closely resembles that of Schwarzschild black holes.

However, as the horizon radius approaches the quantum scale parameter $l$ , the graphs deviates from that behavior.   While free energy of the Schwarzschild blackhole decreases monotonically, the Hayward curve develops a local minimum  and decrease tends the plots to bend upward to higher free energy value. 

This minimum value indicates that, during evaporation , the blackhole reached the most thermodynamically favorable configuration. After this point, the graph bends upward . this indicates that if the blackhole wants to undergo further evaporation, it requires an increase in free energy. Which means, the continuous reduction of radius or evaporation is energetically unfavorable. so the blackhole reaches a stable remnant  instead of complete evaporation. 
Another important feature is, the effect of the regularization parameter $l$ . Larger value of $l$ shifts the minimum point and the remnant to larger horizon radii so as higher energy values. This indicates that stronger the value of quantum correction stops the evaporation earlier and stabilizes the black hole with larger radii. 
Another interesting observation is, the possibility of two different radii supporting same free energy value. as we have already seen this in the case of mass and temperature studies. Let's discuss that in the  next subsection. \\

\textbf{Helmholtz Free Energy vs. Entropy} \\
This figure (\ref{fig:F_vs_s}) provides information on the thermodynamic evolution of the blackhole in terms of entropy. As all other cases, here also we can see Hayward blackhole behaves similar to Schwarzschild blackhole for larger entropy regions. In this region, free energy increases with entropy almost linearly, as classical blackholes behaves.

As the blackhole evaporates, its entropy and free energy decreases. Unlike classical blackholes, instead of evolving to zero free energy state, after a minium free energy the curve bends upward. For larger value of $l$ , this correction takes place for larger entropy. This indicates that, due to quantum correction, the black hole stabilizes before it reaches extremely small entropy. 
when this graph is observed align with temperature entropy graph, it is clear that, the free energy graph reaches its end at lower entropy region where the temperature temperature reaches zero entropy. So it is clear that the blackhole trapped in a cold remnant state instead of a complete evaporation. 

\subsection{Thermodynamic Critical Points}

By analyzing the results, that are presented in the previous sections, we have mentioned about two radii that support same value in the case of mass, temperature and free energy. We can identify the critical radius corresponding to extrema by setting the derivative with  respect to $r_h$ equal to zero.

The mass of the Hayward black hole is given by

\begin{equation}
M(r_h)=\frac{r_h^3}{2(r_h^2-l^2)}.
\end{equation}

Differentiating with respect to the horizon radius yields

\begin{equation}
\frac{dM}{dr_h}
=
\frac{r_h^2(r_h^2-3l^2)}
{2(r_h^2-l^2)^2}.
\end{equation}

The condition

\begin{equation}
\frac{dM}{dr_h}=0
\end{equation}

gives

\begin{equation}
r_h=\sqrt{3}\,l.
\end{equation}

This is the radius corresponding to the minimum mass that can be attained by the blackhole.

For the Hawking temperature,
\begin{equation}
T_H(r_h)
=
\frac{r_h^2-3l^2}
{4\pi r_h^3},
\end{equation}

whose derivative is

\begin{equation}
\frac{dT_H}{dr_h}
=
\frac{9l^2-r_h^2}
{4\pi r_h^4}.
\end{equation}

The temperature reaches its maximum when

\begin{equation}
\frac{dT_H}{dr_h}=0,
\end{equation}

which occurs at

\begin{equation}
r_h=3l.
\end{equation}

The specific heat can be written as

\begin{equation}
C
=
\frac{dM}{dT}
=
\frac{2\pi r_h^6(r_h^2-3l^2)}
{(r_h^2-l^2)^2(9l^2-r_h^2)}.
\end{equation}

As we could see a  discontinuity, in the specific heat graph, It is evident that the numerator vanishes at the extremal radius,
$r_h=\sqrt{3}l$, whereas the denominator vanishes at
$r_h=3l$. Thus, the thermodynamic stability is completely determined by these two characteristic radii.

Further insight is obtained from the Helmholtz free energy,

\begin{equation}
F=M-TS.
\end{equation}

Using the first law of black-hole thermodynamics,

\begin{equation}
dM=T\,dS,
\end{equation}

its differential becomes

\begin{equation}
dF
=
dM-T\,dS-S\,dT
=
-S\,dT.
\end{equation}

Since the entropy remains positive throughout the physical domain,

\begin{equation}
\frac{dF}{dr_h}
=
-S\frac{dT}{dr_h}.
\end{equation}

Since the minimum of the Helmholtz free energy occur at $\frac{dF}{dr_h}=0$, then one can easily say that this is at , 

\begin{equation}
\frac{dT}{dr_h}=0,
\end{equation}

implying

\begin{equation}
r_h^{(F_{\min})}=3l.
\end{equation}

Hence, the minimum of the Helmholtz free energy coincides exactly with the Davies critical point, establishing a direct connection between energetic preference and thermodynamic stability.

These analytical results explain the non-monotonic behavior observed in the mass, temperature and Helmholtz free-energy curves. Between the two characteristic radii,

\begin{equation}
\sqrt{3}\,l<r_h<3l,
\end{equation}

the thermodynamic functions become non-monotonic, allowing two distinct horizon radii to correspond to the same value of mass, Hawking temperature or Helmholtz free energy. These solutions belong to different thermodynamic branches and are separated by the Davies critical point. Such multivalued behavior is absent in the Schwarzschild black hole and arises naturally from the regularizing length scale introduced by the Hayward geometry.

\begin{table}[htbp]
\centering
\caption{Characteristic thermodynamic radii of the Hayward black hole.}
\label{tab:thermo_radii}
\begin{tabular}{c c p{0.48\columnwidth}}
\hline\hline
Radius & Mathematical Condition & Physical Significance \\
\hline
$r_h=\sqrt{3}\,l$ & $\displaystyle \frac{dM}{dr_h}=0$ & Extremal remnant, minimum mass, $T_H=0$, $C=0$ \\ [1.5ex]
$r_h=3l$ & $\displaystyle \frac{dT_H}{dr_h}=0$ & Maximum temperature, Davies critical point, $C\rightarrow\infty$, minimum Helmholtz free energy \\
\hline\hline
\end{tabular}
\end{table}

\subsection{Dual Thermodynamic Branches}

A remarkable feature that emerges consistently from the thermodynamic analysis is the existence of two distinct horizon radii corresponding to the same thermodynamic state. This behaviour is observed in the mass--radius, Hawking temperature--radius, and Helmholtz free energy--radius curves, where a horizontal line drawn within a certain range intersects the Hayward curve at two different values of the horizon radius. Such multivalued behaviour is absent in the Schwarzschild black hole and arises as a direct consequence of the regularization introduced by the Hayward geometry.

The origin of this behaviour can be understood from the non-monotonic nature of the thermodynamic functions. The mass possesses a minimum at

\begin{equation}
r_h=\sqrt{3}\,l,
\end{equation}

while the Hawking temperature reaches a maximum at

\begin{equation}
r_h=3l.
\end{equation}

Between these two characteristic radii,

\begin{equation}
\sqrt{3}\,l<r_h<3l,
\end{equation}

the thermodynamic quantities are no longer one-to-one functions of the horizon radius. Consequently, a single value of the mass, Hawking temperature, or Helmholtz free energy can correspond to two different black-hole configurations.

These two solutions represent distinct thermodynamic branches. The branch with the smaller horizon radius lies between the extremal remnant and the Davies critical point, where the specific heat is positive. This branch is therefore locally thermodynamically stable and corresponds to the quantum-corrected remnant phase. The second branch, associated with larger horizon radii, possesses negative specific heat and behaves similarly to the classical Schwarzschild black hole, representing the unstable evaporating phase.

The Davies critical point separates these two branches. At this point the specific heat diverges, the Hawking temperature reaches its maximum value, and the Helmholtz free energy becomes minimum. Therefore, the dual-branch structure is not merely a geometrical feature of the thermodynamic curves, but a direct manifestation of the transition between stable and unstable thermodynamic phases.

From a physical perspective, the existence of two horizon radii for the same thermodynamic state reflects the regular nature of the Hayward geometry. During evaporation, the outer event horizon continuously shrinks while the inner horizon expands. Both horizons gradually approach one another until they merge at the extremal configuration,
\begin{equation}
r_h=\sqrt{3}\,l,
\end{equation}
where the Hawking temperature vanishes and the evaporation process terminates with the formation of a stable remnant. Thus, the observed dual-branch behaviour provides a thermodynamic signature of the regular black-hole structure and distinguishes the Hayward solution from the classical Schwarzschild spacetime.
\subsection{Unified Thermodynamic State Space}

\begin{figure}
    \centering
    \includegraphics[width=0.7\linewidth]{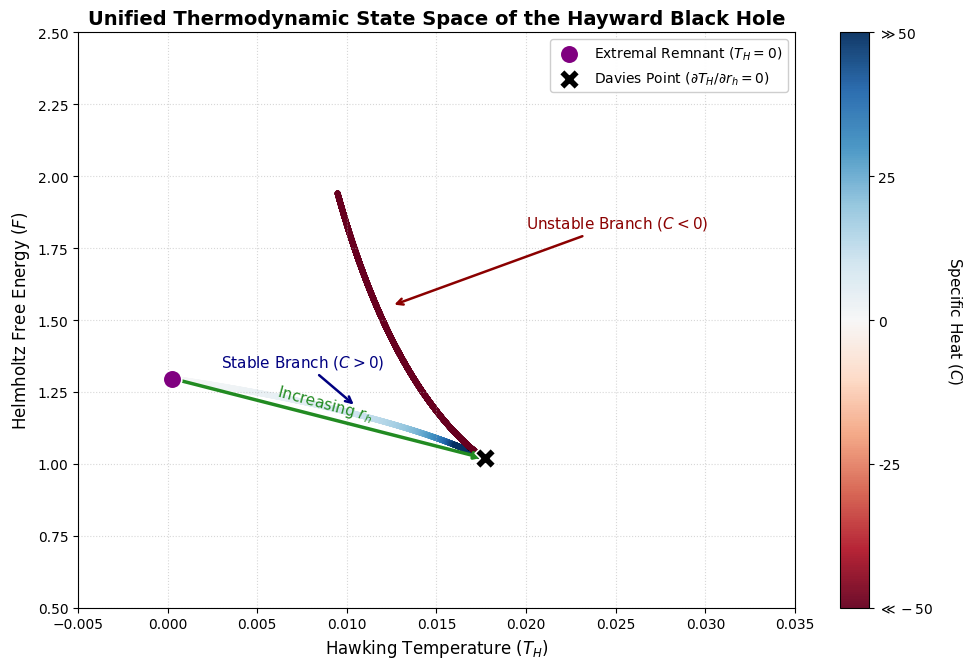}
    \caption{This figure show a combined picture of how the effect of specific heat, free energy and Hawking temperature influences the evolution of the Hayward blackhole as radius changes.}
    \label{fig:state_space}
\end{figure}

Figure~\ref{fig:state_space} presents a unified thermodynamic state-space representation of the Hayward black hole by simultaneously combining the Helmholtz free energy, Hawking temperature, and specific heat. In this representation, the horizontal axis shows the Hawking temperature, the vertical axis represents the Helmholtz free energy, and the color scale indicates the specific heat.

We can begin the evolution at the extremal remnant state, which is marked by the purple circle, where the Hawking temperature vanishes,
\begin{equation}
T_H=0.
\end{equation}
we have already this point, corresponds analytically to the extremal radius
\begin{equation}
r_h=\sqrt{3}\,l,
\end{equation}
at which the mass reaches its minimum value,
\begin{equation}
\frac{dM}{dr_h}=0,
\end{equation}
and the specific heat also vanishes,
\begin{equation}
C=0.
\end{equation}
The remnant therefore represents a thermodynamically inert endpoint of the evaporation process.

If the horizon radius increased beyond the extremal configuration, the Hawking temperature also increases but the Helmholtz free energy decreases. During this stage the specific heat remains positive, as indicated by the blue colour map, showing that the black hole occupies a locally stable thermodynamic branch. This same  behavior can be observed with the specific-heat plots presented previously, where the interval
\begin{equation}
\sqrt{3}\,l<r_h<3l
\end{equation}
can be identified as the stable branch connecting the extremal remnant to the Davies critical point.

The evolution terminates at the Davies critical point, which is represented by the black cross. Analytically, this point is,
\begin{equation}
\frac{dT_H}{dr_h}=0,
\end{equation}
which we already found as,
\begin{equation}
r_h=3l.
\end{equation}
At this radius the Hawking temperature reaches its maximum value, while the Helmholtz free energy attains its minimum. This follows directly from the first law of black-hole thermodynamics,
\begin{equation}
dF=-S\,dT,
\end{equation}
which implies
\begin{equation}
\frac{dF}{dr_h}
=
-S\frac{dT}{dr_h}.
\end{equation}
Since the entropy remains positive throughout the physical region, the extrema of the Helmholtz free energy coincide with the extrema of the Hawking temperature. Consequently, the minimum of the free-energy curve occurs precisely at the Davies radius.

At the same point, the specific heat diverges,
\begin{equation}
C\rightarrow\pm\infty,
\end{equation}
This shows  the transition between thermodynamic branches. The sudden change in the colour map from blue to red indicates this change of stability, corresponding respectively to positive and negative specific heat. Thus, the Davies point represents the boundary separating the locally stable and unstable thermodynamic phases of the Hayward black hole.

The unified state-space diagram therefore summarizes the complete thermodynamic evolution obtained from the individual mass, temperature, entropy, specific-heat, and free-energy analyses. The extremal radius,
$r_h=\sqrt{3}l$, marks the formation of a stable remnant with vanishing temperature, whereas the Davies radius,
$r_h=3l$, identifies the thermodynamic critical point where the free energy is minimized and the stability changes. Together, these two characteristic radii govern the thermodynamic behavior of the Hayward black hole and provide a unified interpretation of all the thermodynamic quantities considered in this work.

\section{Conclusion}
In this work, we could make a Thermodynamic investigation of the Hayward black hole, in comparison with that of Schwarzschild blackhole. We could uncover the evolution of the blackhole by analyzing the change in mass, Hawking temperature, specific heat and Helmholtz Free energy. It was evident that the regularization parameter $l$ could make significant deviation from the behavior of classical blackholes near the quantum regime. but its thermodynamic behavior for large radii stay similar to that of classical black holes.

The analysis shows that the thermodynamic evolution of the Hayward black hole is governed by two characteristic radii. The extremal radius,
\[
r_h=\sqrt{3}\,l,
\]
corresponds to the minimum mass configuration, where the Hawking temperature and specific heat reaches zero, indicating the formation of a stable cold remnant that stops the evaporation process. \\
The second characteristic radius,
\[
r_h=3l,
\]
defines the Davies critical point, where the Hawking temperature reaches its maximum, the Helmholtz free energy reaches minimum, and the specific heat diverges, marking the transition between locally stable and unstable thermodynamic branches.

 Between these two critical radii, multiple horizon radii can correspond to the same thermodynamic quantities, indicating the existence of distinct thermodynamic branches that arise naturally from the regularized geometry. This type of  behavior is absent in the Schwarzschild solution and it shows the role of quantum corrections in modifying black-hole thermodynamics.

To summarize the complete thermodynamic evolution, we introduced a unified thermodynamic state-space representation that combines the Helmholtz free energy, Hawking temperature, and specific heat in a single diagram. This representation provides a visualization of the evolution from the extremal remnant to the Davies critical point and shows the relationship between preferred energy states and thermodynamic stability.

The present study demonstrates that the Hayward regular black hole possesses a rich thermodynamic structure characterized by remnant formation, thermodynamic phase transitions, and stability changes as due to the  regularization parameter. These results contribute to the understanding of regular black-hole thermodynamics and may provide useful insights into quantum-gravity motivated models that we can use to  resolve spacetime singularities. Also the non-radiating zero temperature core with mass can be a possible candidate for dark matter. 
Future investigations may extend this analysis to rotating or charged regular black holes, explore the effects of a cosmological constant or extended phase-space thermodynamics, and examine possible observational consequences of remnant formation and thermodynamic critical behavior.

\section*{Acknowledgments}
The authors expresses gratitude to the anonymous reviewers for comments that improved the manuscript.

\section*{Declarations}

\subsection*{Funding}
The author declared that no grants, funds, or other financial support were received for the preparation of this manuscript. 

\subsection*{Competing Interests}
The author has no financial or proprietary interests in any material discussed in this article.

\subsection*{Data Availability Statement}
The data supporting the findings of this study (including analytical models and generated data plots) are available within the article or can be provided by the corresponding author upon reasonable request.

\subsection*{Author Contributions}
The entirely of this work, including the analytical derivations, physical interpretations, numerical plotting, and manuscript composition, was performed solely by the authors.

\bibliography{References}
\end{document}